\documentclass[a4paper,11pt]{article}
\pdfoutput=1 % if your are submitting a pdflatex (i.e. if you have
\usepackage{jinstpub} % for details on the use of the package, please
\usepackage{xcolor}
\usepackage{lineno}
\title{\boldmath Scintillation characteristics of a NaI(Tl) crystal at low-temperature with silicon photomultiplier
}

\author[a]{H. Y. Lee}
\author[a]{J. A. Jeon}
\author[a]{K. W. Kim}
\author[b,a]{W. K. Kim}
\author[a,b]{H. S. Lee}
\author[a,b]{M. H. Lee}

\affiliation[a]{Center for Underground Physics, Institute for Basic Science (IBS),\\Daejeon 34126, South Korea}
\affiliation[b]{University of Science and Technology (UST),\\ Daejeon 34113, South Korea}

\emailAdd{kwkim@ibs.re.kr}
\emailAdd{hyunsulee@ibs.re.kr}

\abstract{
	Scintillation characteristics of a thallium doped sodium iodide (NaI(Tl)) crystal with a dimension of 0.6 $\times$ 0.6 $\times$ 2\,$cm^{3}$ are studied by attaching a silicon photomultiplier (SiPM) direct to the crystal  over a temperature range from 93 to 300\,K. 
    The scintillation light output and decay time are measured by irradiating 59.54\,keV $\gamma$-rays from a $^{241}$Am source. 
	 	We observed approximately 20\% increase in light yield at 230\,K compared to that at the room temperature.
	  At this condition, the NaI(Tl) crystal coupled with the SiPM can be a good candidate for future dark matter search	detector. 

}

\begin{document}
\maketitle
\flushbottom

\section{Introduction}
\label{sec:intro}

Due to excellent scintillation properties, NaI(Tl) crystals are widely used to measure energies of gammas from environmental radioactivities~\cite{doi:10.1146/annurev-matsci-070616-124247} and from interactions of new physics particles~\cite{Kim_2010,Adhikari:2018ljm,Adhikari:2019off,KIMS:2018hch,Amare:2019jul,Amare:2021yyu}. 
In particular, a direct detection of dark matter particles using NaI(Tl) crystal detectors is currently being conducted in several experiments~\cite{Adhikari:2017esn,Amare:2018sxx,sabre,Bernabei:2020mon}.
The low-background NaI(Tl) crystals have been developed for a success of the dark matter detection at low energies by reducing internal radioactive contaminants in the crystal~\cite{Shin:2018ioq,Suerfu:2019snq,Park:2020fsq,Fushimi:2021mez}. 

Conventional NaI(Tl) detectors are operating at the room temperature with photomultiplier tubes (PMTs) attached at each end of the crystal~\cite{Bernabei:2012zzb,Adhikari:2017esn}. 
On the other hands, PMTs are bulky and contain relatively large amount of radioisotopes~\cite{cosinebg,Adhikari:2021rdm}. 
The photodetection efficiency (PDE) of the recently developed PMT, R12669SEL from Hamamatsu photonics, has the maximum value of 40\% at around 420\,nm corresponding to the emission peak of scintillation lights from the NaI(Tl) crystals. 
Silicon photomultipliers (SiPMs) could be good replacements of the PMTs because of their high PDEs for a broad range of wavelength and compact size with less radioactive materials~\cite{Baudis:2018pdv}. 
A high dark count rate (DCR) of the SiPM was a contemporary problem for the application of the low-energy dark matter search experiment. 
However, DCR of the SiPM rapidly decrease at low temperatures~\cite{Lightfoot:2008im}.  A few experiments are conducting rare event searches using the SiPM already~\cite{nEXO:2017nam,DarkSide-20k:2017zyg}. 

Nonlinear temperature dependence of the NaI(Tl) crystals' scintillation light yield had been observed by a few measurements with PMT readout~\cite{IANAKIEV2009432,Sailer:2012ua}. 
Especially, increased light yields at temperatures around 150 and 250\,K compared to that at the room temperature were reported~\cite{Sailer:2012ua}. 
In this case, the NaI(Tl) crystal coupled with the  SiPM photosensor at the low temperatures can be an ideal combination for the dark matter search experiments.
In this paper, we report responses of the NaI(Tl) crystal coupled with the SiPM sensor in a temperature range of 93 $-$ 300\,K and explore a possibility for the future dark matter search experiment. 

%properties of the NaI(Tl) crystal using a SiPM readout are obtained for various temperature ranges and compared with the previously reported results. %In addition, three SiPM sensors with different window conditions were used to check whether the absolute light yield of the crystal was changed.

\section{Experimental setup}
\label{sec:exp}
We use a NaI(Tl) crystal with a dimension of 0.6\,cm $\times$ 0.6\,cm $\times$ 2\,cm that was cut from a crystal ingot grown in a program for the low-background NaI(Tl) development reported as NaI-036 in Ref.~\cite{Park:2020fsq}. 
Several layers of soft polytetrafluoroethylene (PTFE) sheets were wrapped on side five faces of the crystal. 
The SiPMs from Hamamatsu photonics, model number S13360-6075CS with an active area of 0.6\,cm $\times$ 0.6\,cm, were attached to one side of the crystal. 
A micropixel size of 75\,$\mu$m $\times$ 75\,$\mu$m, a total of 6,400 micropixels, and a 50\% PDE at 450\,nm were provided in the specification by the company. 
%An active area of 6\,mm $\times$ 6\,mm with a micropixel size of 75\,$\mu$m, a total 6400 micropixels, and a 50\% photodetection efficiency (PDE) at 450\,nm wavelength are company specifications of the device.

\begin{figure}[htbp]
\centering % \begin{center}/\end{center} takes some additional vertical space
\includegraphics[width=.9\textwidth]{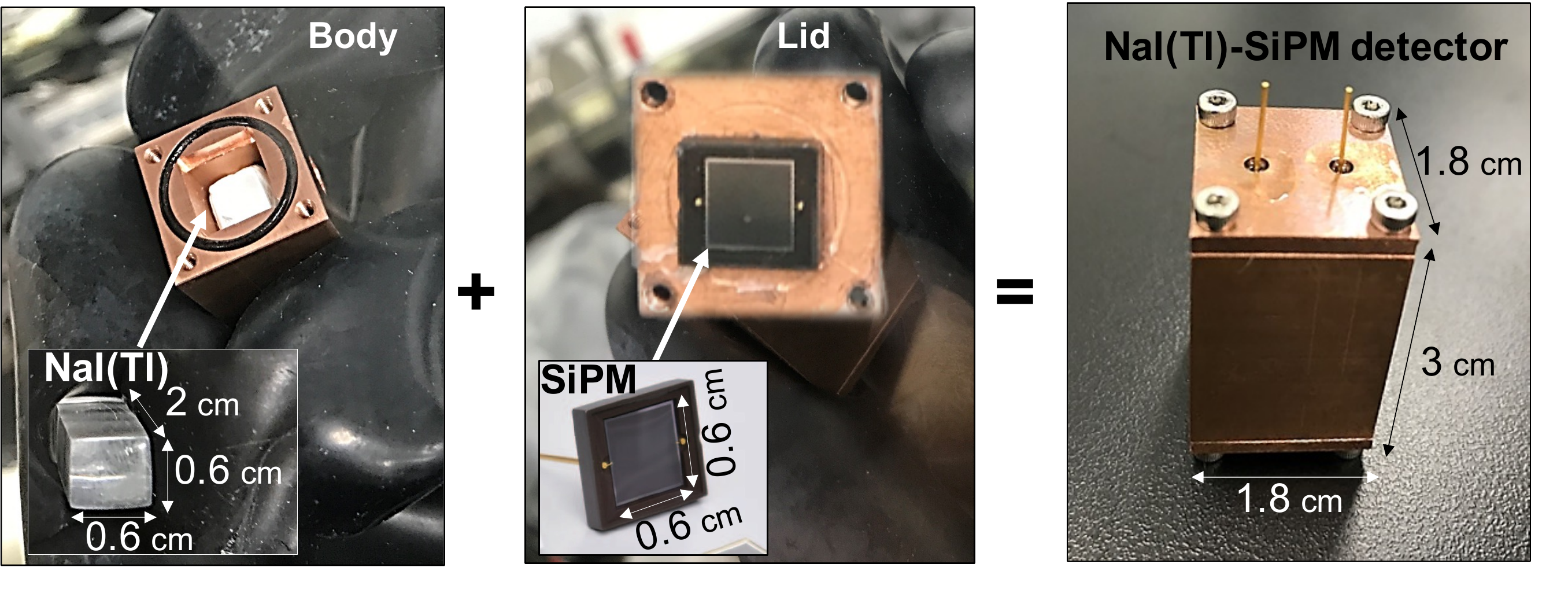} \\
(a) \\
\includegraphics[width=.9\textwidth]{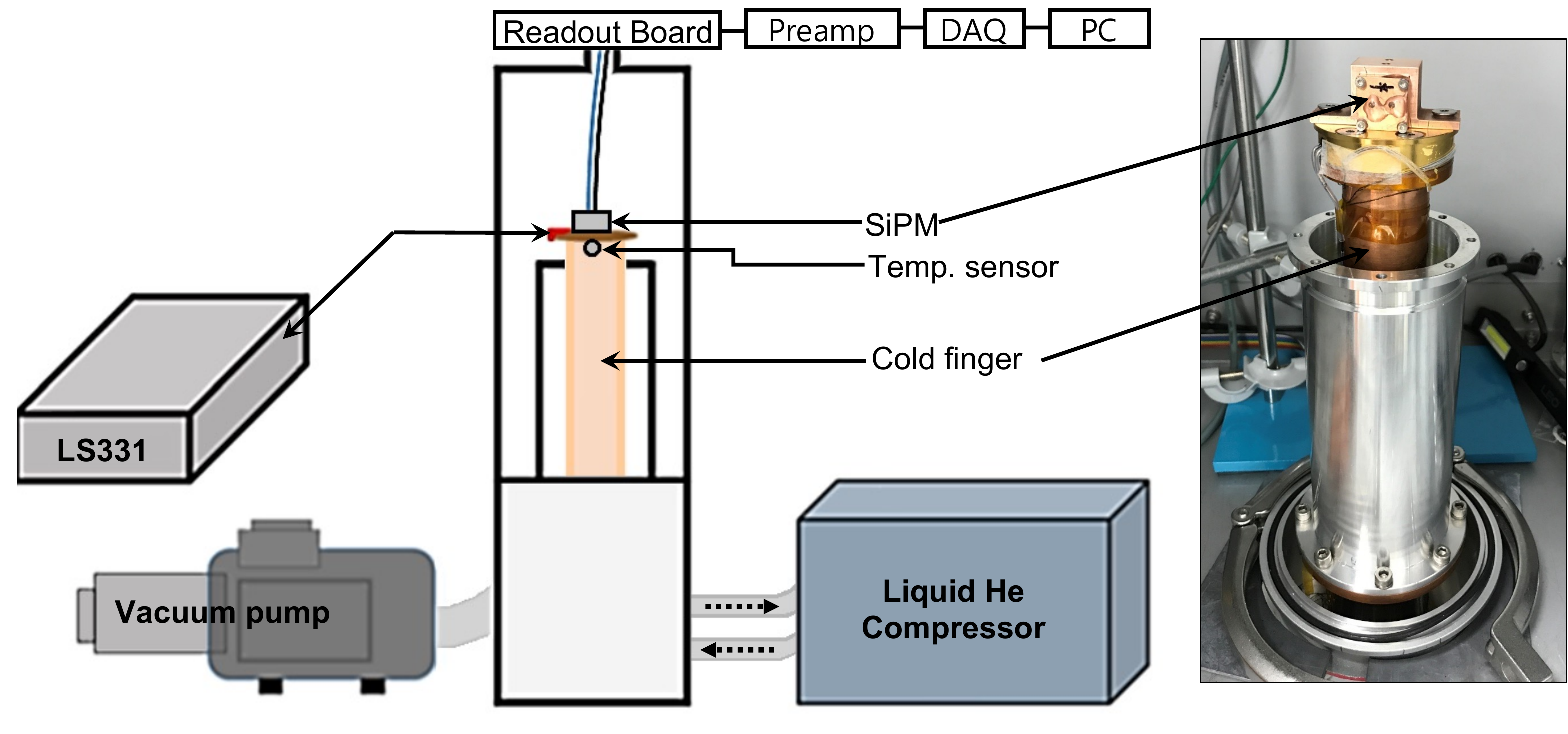} \\
(b) \\
\caption{\label{fig:setup} (a) NaI(Tl)-SiPM prototype assembly photos (b) Experimental setup for low-temperature measurements}
\end{figure}

The NaI(Tl) crystal was encapsulated in a copper housing to avoid moisture in the air due to its highly hygroscopic nature. The SiPM was coupled to the front face of the crystal in the housing named as "NaI(Tl)-SiPM prototype" as shown in Fig.~\ref{fig:setup} (a). 
The crystal was inserted into the housing and the SiPM was attached to the lid of the housing. 
The crystal and the SiPM had approximately 2\,mm gap to maintain the same optical coupling condition. 
A Viton o-ring between the lid and the box was pressed with screws for an air tighteness of the NaI(Tl) crystal. 
All assembly processes were done inside a glove box in which the humidity level was maintained to be less than a few tens of ppm using N$_2$ gas and a molecular sieve trap. 
The dimension of the NaI(Tl)-SiPM prototype was 1.8\,cm $\times$ 1.8\,cm $\times$ 3\,cm (W, D, H) as shown in Fig.~\ref{fig:setup} (a). 
A $^{241}$Am source was placed above the copper housing to irradiate 59.54\,keV $\gamma$-rays to the crystal.
%Before the capsule assembly, the crystal was wrapped in several layers of Tetratex sheets to prevent light loss. 

%We have about 2\,mm gab between the SiPM and the crystal to avoid an impact due to changes of optical coupling condition caused by different thermal expansion coefficient. 

The NaI(Tl)-SiPM setup was installed in a cryostat that could control the temperature from 4\,K to the room temperature~\cite{Pandey:2020cbz}.
The schematic diagram of the cryostat is shown in Fig.~\ref{fig:setup} (b). A cold finger was cooled by a liquid helium compressor.  The temperature of the cold finger was monitored by a silicon diode sensor and controlled through the Lakeshore LS331 temperature controller. % to the heater. 
%To avoid heat loss of the cold finger and the NaI(Tl)-SiPM prototype, a thermal cap shield was installed. 
The vacuum was maintained at approximately 10$^{-2}$ Torr by a rotary pump. 
The NaI(Tl)-SiPM prototype was placed inside the thermal shield and directly attached  to the cold finger. 
On the top of the NaI(Tl)-SiPM prototype, we installed one another temperature sensor. 
The detector was tested in a temperature range between 93 and 300\,K of the cold finger. 
The experiment was performed at each temperature point after a few hours of waiting on the demanded temperature considering heat transfer to the crystal. 
Temperatures of the cold finger and the top of the NaI(Tl)-SiPM prototype were continuously measured and found  about 3-5\,K differences between two points that were accounted as a systematic uncertainty. 
%We take the temperature measured from the sensor attached to the cold finger as a crystal temperature in this study. But there was a persistent temperature difference of 3-5 K between the cold finger and top of the NaI(Tl)-SiPM detector copper jig. This difference was considered as an uncertainty for the temperature.
The SiPM readout board was installed outside the cryostat. 
An amplified signal from the SiPM readout was digitized by a 125\,MHz, 12-bit flash-analog-to-digital-converter (FADC). 
%Considering scintillation decay times of the NaI(Tl) crystal depending on temperature, 
A 32\,$\mu$s-long waveform was stored for each triggered event.
Three different window types of resin, quartz, and windowless were tested for this measurement.
%e SiPM sensors were tested to confirm whether the absolute light yield of the crystal measured with the different window conditions was different. The model of the SiPMs is all same, but they are just different the window conditions. There are Si resin, quartz, and windowless conditions.

%\section{Results on the characterizations of the NaI(Tl)-SiPM detector}
\section{SiPM only measurements}
\label{sec:res}
A SiPM is a pixelated device where each micropixel is a series of an avalanche photodiode (APD) and a quenching resistor. 
The SiPM is externally biased with bias voltage $V_{bias}$ so that the voltage on each APD is above its breakdown voltage $V_{BD}$. 
The difference $V_{bias} - V_{BD}$ is known as overvoltage $\Delta$V which is one of the most important parameters affecting the performance of the SiPM. 
%SiPM is operated at a bias voltage ($V_{bias}$) higher than the breakdown voltage . $V_{bias}$ exceeds $V_{BD}$ by an overvoltage $\Delta$V ($V_{bias}$ -$V_{BD}$), which has a significant impact on detector performance. 

\begin{figure}[htbp]
\centering % \begin{center}/\end{center} takes some additional vertical space
\includegraphics[width=.7\textwidth]{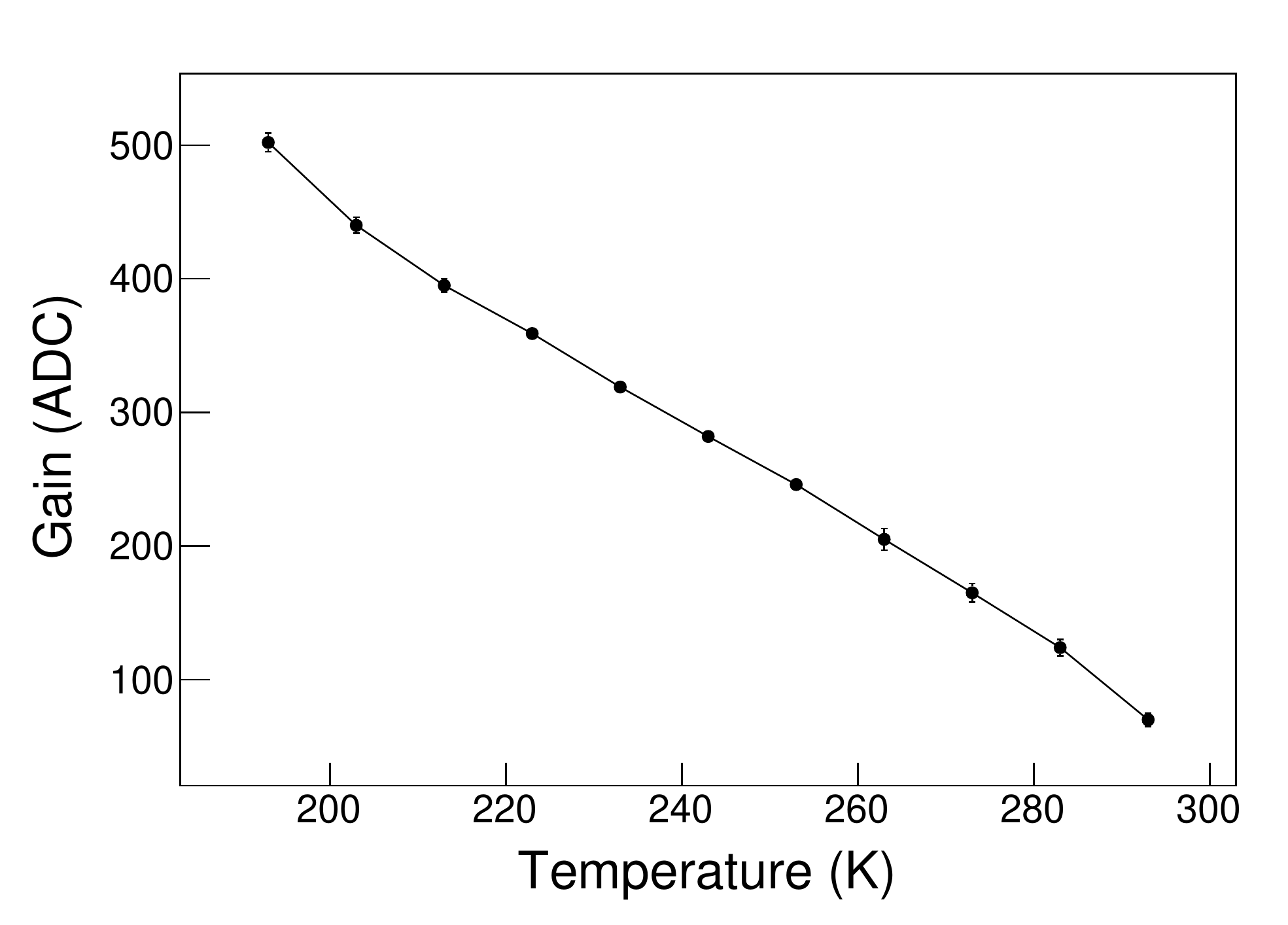} 
\caption{\label{fig:gain}The SPE gain as a function of temperature with same $V_{bias}$ of 54\,V.  }
\end{figure}

To understand the characteristics of the SiPM, we first tested the SiPM without NaI(Tl) crystal in the cryostat. 
The $V_{BD}$ was decreased at low temperatures~\cite{Baudis:2018pdv,COLLAZUOL2011389} because of ionization rate increase~\cite{sze2021physics}. The results of increasing gain in the low temperature with the same $V_{bias}$ are shown in Fig.~\ref{fig:gain}. 
To avoid an impact caused by large gain changes from the temperature variation, we maintain $\Delta$V equal to 3\,V in the following measurements. 
Typically a single photoelectron (SPE) signal has a height of 12\,mV and a width of 100\,ns in this condition. 

%Here we apply $V_{bias}$ equal to 54\,V and obtain the gain in a range of temperature between 200\,K and 300\,K. Linear increase with temperature decreasing is clearly observed. 
%For the following measurements, we apply the bias voltage to have a similar gain in different temperatures by maintaining $\Delta$V equal to 3\,V. 
%The amplitude and recovery time of a single photoelectron (SPE) of the SiPM by the readout board is about 12\,mV and about 100\,ns respectively. 
%In Fig.~\ref{fig:gain-dcr} (a), the gain of the SiPM at 54 V linearly increased in the temperature range of 200 K < T < 300 K. For this reason, in the NaI(Tl)-SiPM detector experiment at low temperature, the change in $V_{BD}$ as a function of temperature was reflected in order to keep the constant operating conditions of the SiPM (the same gain).

\begin{figure}[htbp]
\centering % \begin{center}/\end{center} takes some additional vertical space
\includegraphics[width=.8\textwidth]{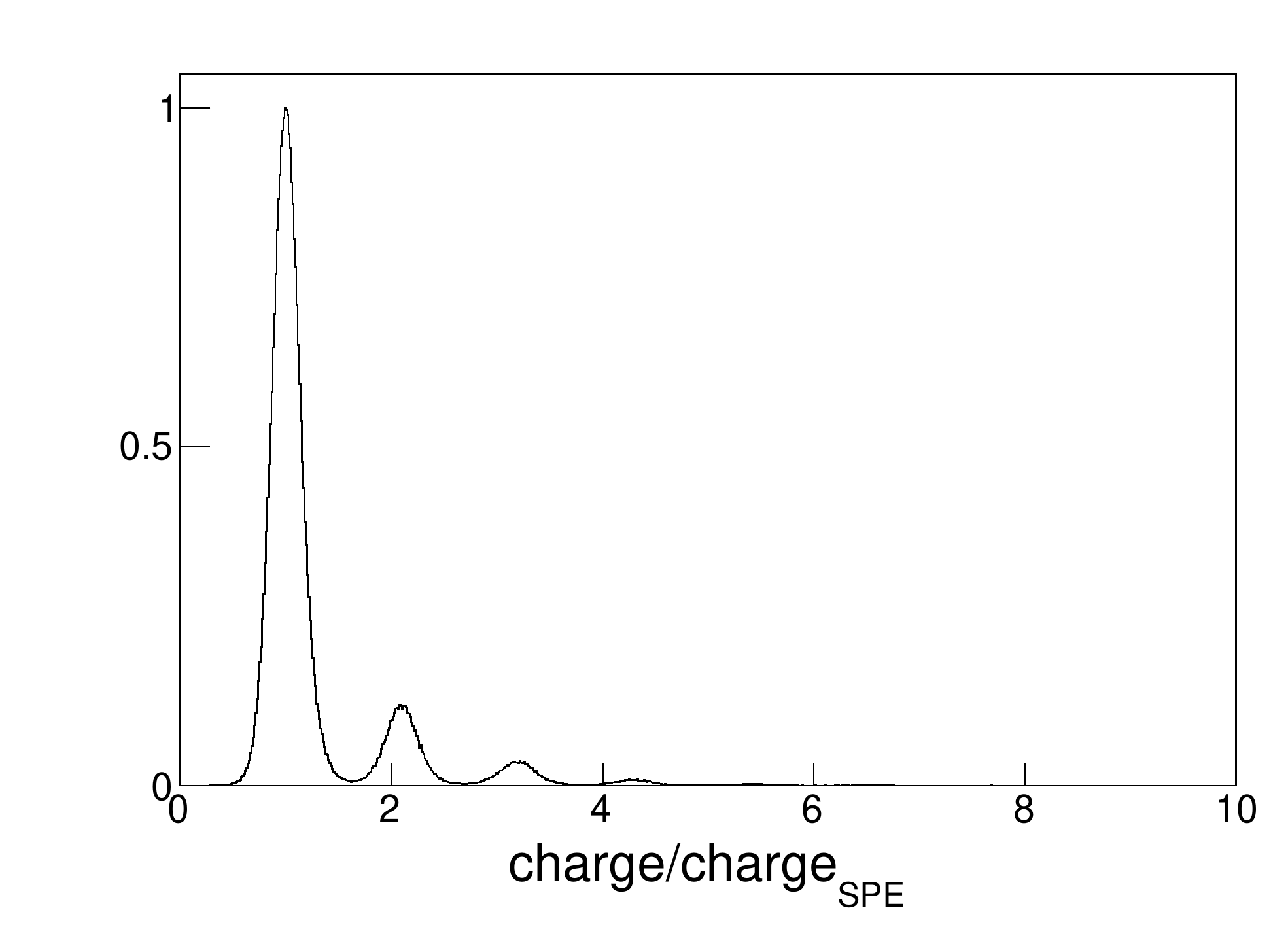}
\caption{\label{fig:pecharge} Charge distribution of the dark count events at the room temperature is presented. The typical dark current events provide a clear peak at single photoelectron but there are multi-photoelectrons events above 1.5 single photoelectron charge caused by cross talks. 
}
\end{figure}

\begin{figure}[htbp]
\centering % \begin{center}/\end{center} takes some additional vertical space
\includegraphics[width=.7\textwidth]{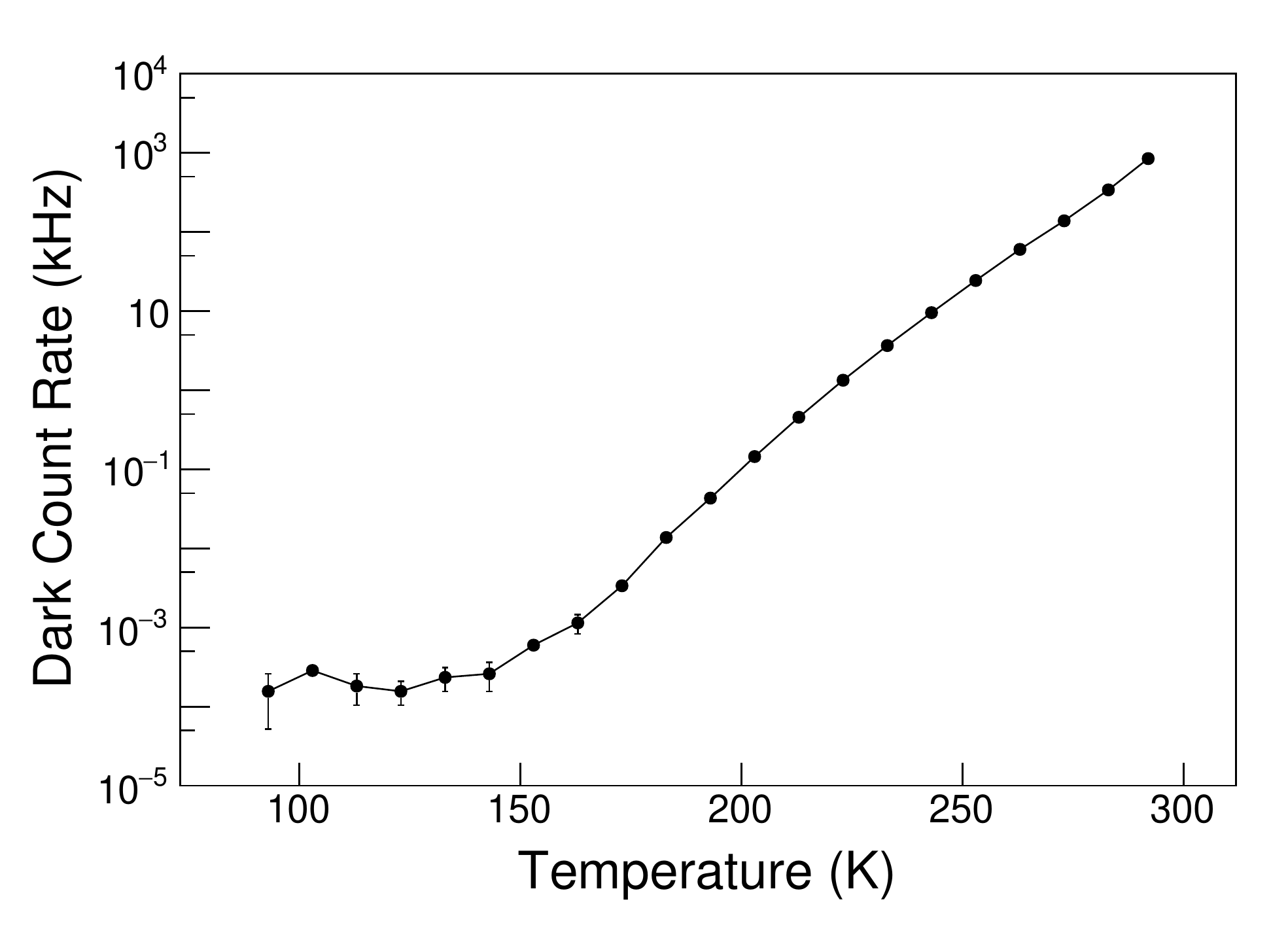} 
% "\includegraphics" from the "graphicx" permits to crop (trim+clip)
% and rotate (angle) and image (and much more)
\caption{\label{fig:dcr}Dark count rates (DCR) in a temperature range 90--300\,K while maintaining $\Delta$V = 3\,V. }
\end{figure}

The SiPM exhibits several noise characteristics such as thermally generated DCR and pixel correlated cross talks. 
By maintaining $\Delta$V equal to 3\,V, we measure the DCR in a temperature range 90--300\,K and an example charge distribution for that at the room temperature is shown in Fig.~\ref{fig:pecharge}. A dramatic decrease of the DCR at low temperatures due to lower thermal generation probabilities~\cite{Otte:2016aaw} is shown in Fig.~\ref{fig:dcr}. 
%A dark count rate (DCR) in the SiPM 
%A thermally generated charge carrier can enter the high field region of an APD and trigger Geiger discharge that is known as a dark count. 
%Because the thermal generation probability is decreased with the temperature decreasing, one can obtain a dramatic decrease of the dark count rate (DCR)~\cite{Otte:2016aaw} as one can see in Fig.~\ref{fig:gain-dcr} (b). 
%This measurement uses the SiPM only setup with $\Delta$V equal to 3\,V. 
From 300 to 140\,K, the DCR decreases six orders of magnitude almost following an exponential function predicted by the Shockley-Read-Hall model~\cite{PhysRev.87.835}. 
DCR is about the same below 140\,K that may be caused by band to band tunneling generation reported in Ref.~\cite{FAcerbi:sipm}. 
%Operation $V_{bias}$ and DCR depending on temperature are later to analyze data of the NaI(Tl)-SiPM setup. 
%The main source of the DCR in this range is expected to thermal generation by field enhancement. But at temperatures lower than 140 K, the DCR wasn’t decreased anymore by the band to band tunneling generation. The values of the DCR at each temperature were reflected in the data analysis of the NaI(Tl)-SiPM detector test. 

Above 1.5 SPE charge accounts  the rate of cross talks that generate multi-photoelectrons events as one can see in Fig.~\ref{fig:pecharge}.
When we evaluate the light yield of the NaI(Tl) crystal, the cross-talk rate is corrected to calculate proper number of photoelectrons. 
%culated from Fig.~\ref{fig:pecharge} is used to correct the light yield of the NaI(Tl) crystal.
%We account the rate of the cross-talk for a calculation of light yields in Fig.~\ref{fig:pecharge}. 
The cross-talk rate is known to be independent from the temperature with the same $\Delta V$~\cite{Baudis:2018pdv,Lightfoot:2008im}, and we also observe a similar result.

\section{NaI(Tl)-SiPM  measurements}
\begin{figure}[htbp]
\centering % \begin{center}/\end{center} takes some additional vertical space
\includegraphics[width=.8\textwidth]{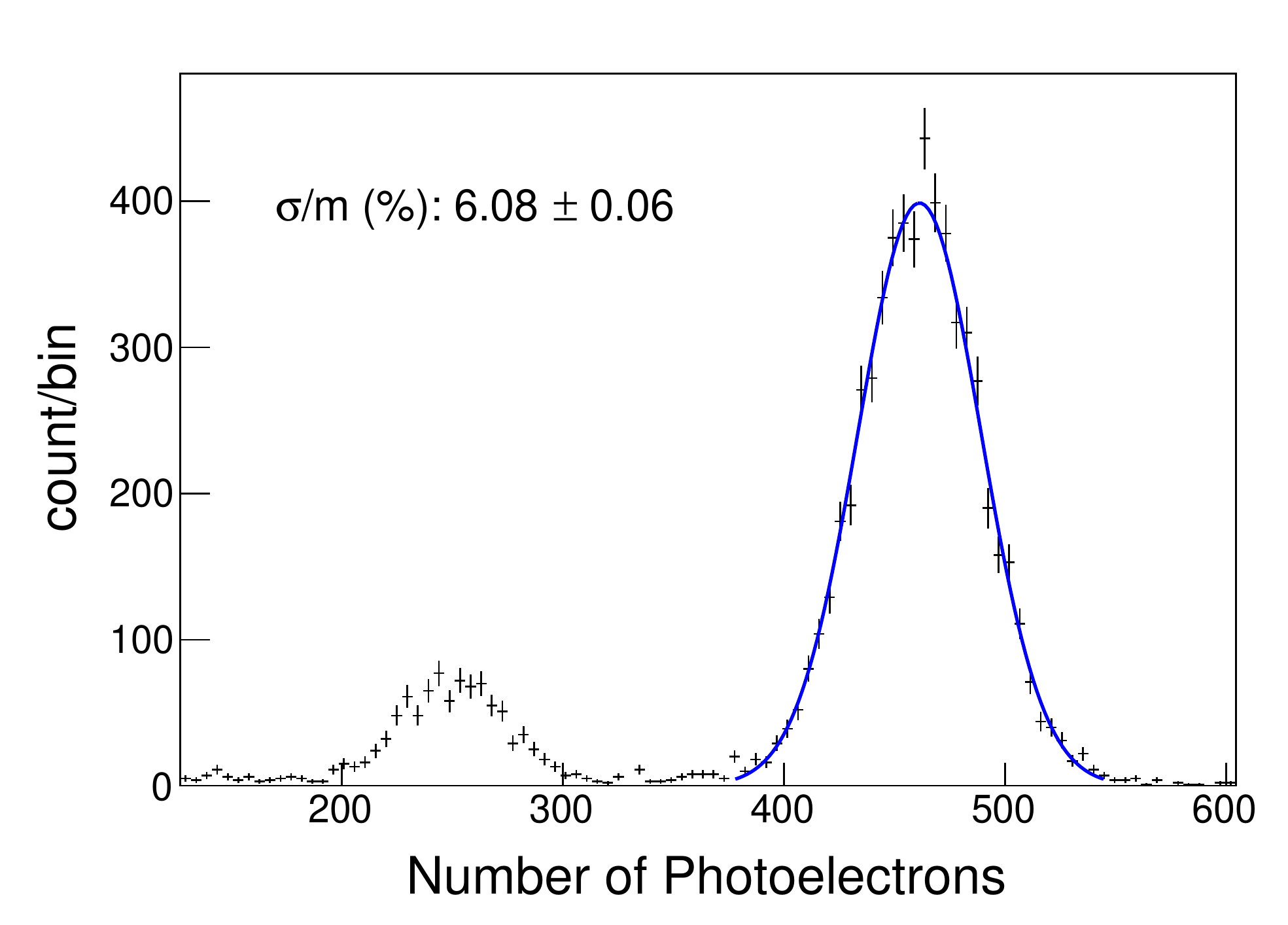}
\caption{\label{fig:am241} The measured NPE with the $^{241}$Am source irradiating NaI(Tl) at room temperature is presented. The peak corresponding to 59.54 keV $\gamma$ is fitted with a Gaussian function (blue solid line). }
%7.75+/-0.03 pe/kev, 6.05+/-0.06 percent sigma/mean
\end{figure}

After a characterization of the SiPM, we configured the NaI(Tl)-SiPM setup to measure light yields of the NaI(Tl) crystals at low temperatures. 
We irradiated 59.54\,keV $\gamma$-rays from a $^{241}$Am source and estimate a measured numbers of photoelectrons (NPE). The measured single photoelectron charge distribution in Fig.~\ref{fig:pecharge} is used to evaluate the light yield per keV together with the correction of the cross-talk rate. 
%A charge distribution of the SPE candidate events is obtained by looking for isolated clusters in the prerecorded pulses of the events as one can see in Fig.~\ref{fig:pecharge}. 
%From the NaI(Tl)-SiPM setup, we measure the temperature dependence of a light yield from 59.54\,keV $\gamma$-ray peak that is divided with a mean charge of the SPE distribution. 
%Light yield is estimated by dividing the total charge deposited by 59.54 keV $\gamma$ into the SPE charge. 
A contribution from DCR is also corrected by the temperature dependence measurement shown in Fig.~\ref{fig:dcr}. 
%The crosstalk is cogenerating the multi-photoelectrons signals are also accounted in the calculation of the light yield although it is not depending on the temperature with same $\Delta$V as shown in Refs.~\cite{Baudis:2018pdv,Lightfoot:2008im}. 
%The charge of 59.54 keV $\gamma$ peak is integrated into the total timing window from a triggered position.
%At room temperature, the DCR is a few hundred kHz which is much higher than that of at low temperatures having below 1 Hz as shown in Fig.~\ref{fig:gain-dcr} (b).
%Therefore the estimated dark count in the timing window which may be accounted as a part of charge from the scintillation was subtracted using the DCR measured independently at each temperature in the total charge calculation.
The distribution of NPE at the room temperature from the $^{241}$Am source is shown in Fig~\ref{fig:am241}. The root-mean-square resolution~($\sigma/m$) is obtained to 6.08$\pm$0.06\,\%. 
%{\color{red} Let's put Am241 data with number of photoelectrons unit (Fit with gaussian and report energy resolution).}.

\begin{figure}[htbp]
\centering % \begin{center}/\end{center} takes some additional vertical space
\includegraphics[width=.8\textwidth]{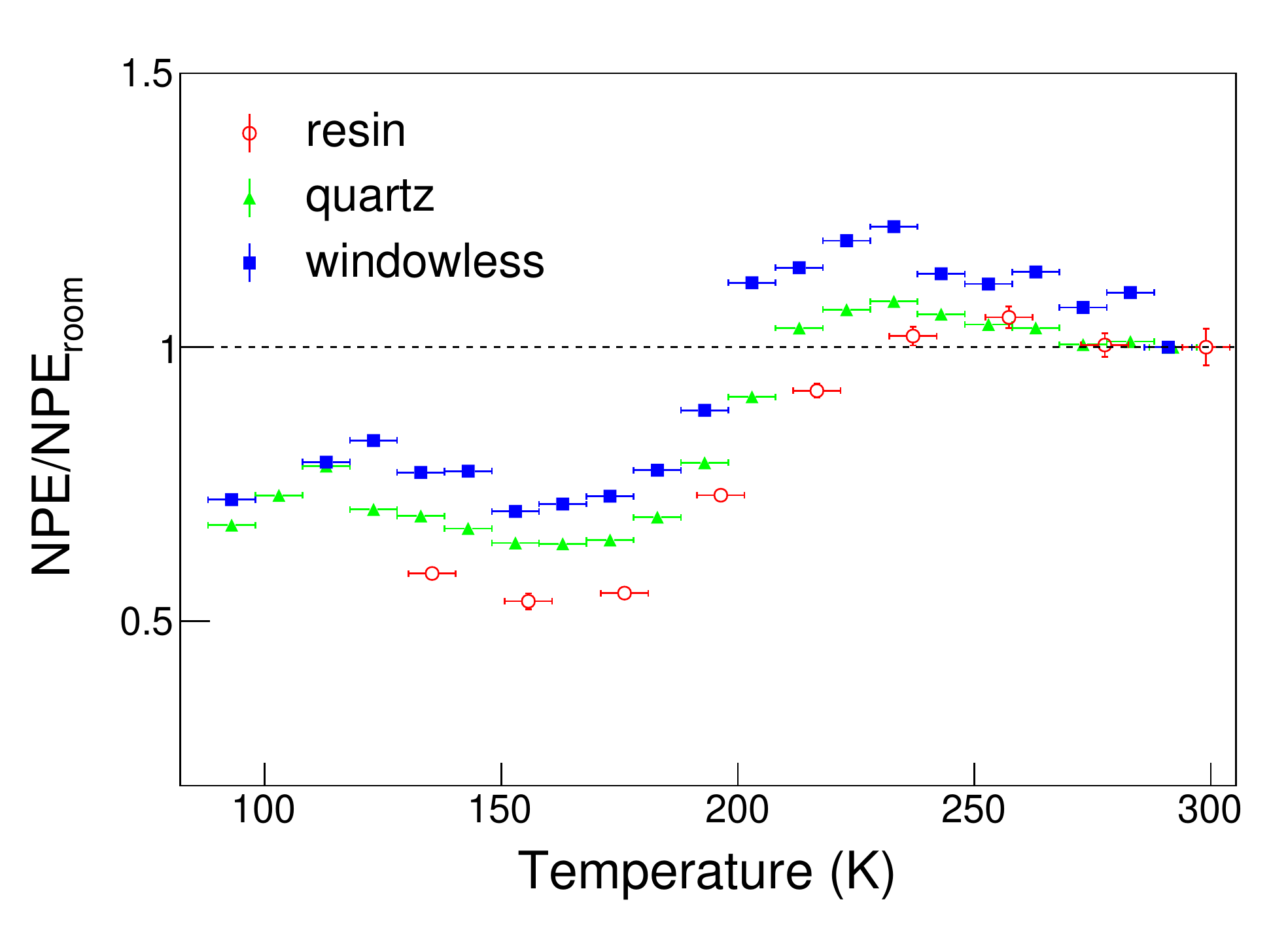}
\caption{\label{fig:npe} The measured NPEs as a function of temperature normalized to that at the room temperature (NPE$_{\mathrm{room}}$) are presented for three SiPMs with different widow types. }
\end{figure}

%Since SiPM has a good resolution to the SPE signal, the SPE charge is estimated by a gaussian fit of the charge distribution of photoelectrons. 
%{\color{red} what about to put SPE and crosstalks distributions plot}.
%However, afterpulse and crosstalk generate multiphoton-like signals that should be accounted for in the evaluation of the SPE charge.
%The SPE charge and crosstalks are shown in Fig.~\ref{fig:pecharge} as the highest peak and multiple integer charges of SPE, representing the probability of the crosstalk.
%But the probability is known to be independent of the temperatures~\cite{Lightfoot:2008im}, so it was not considered in the estimation of the relative light yield between temperatures.
%Charge contribution from the dark count to the total charge of 59.54\,keV also accounts for the multiphoton-like signals. 
%However, there can be generated multi-photons due to the crosstalk effect if the gain is high and DCR is increasing which depends on the temperatures.
%In case of high DCR condition, multi-photon ratio is increasing and can be included in the 59.54 keV signal event which leads to an over-estimated charge of the event.
%We considered this effect by averaging all charges of the multi-photon in the SPE charge estimation.
Figure~\ref{fig:npe} shows NPE as a function of the temperature relative to that of the room temperature (NPE$_{\mathrm{room}}$).
Three different SiPM windows are used for the measurements and show similar dependence of the light yield on temperature. 
Similar results using the PMT readout were reported previously~\cite{Sailer:2012ua}. 
The increased light yields of about 10--20\,\% at 220--250\,K are consistent with Refs.~\cite{Sailer:2012ua,IANAKIEV2009432}. 
However, about 10\% increased light yield at around 150\,K using the PMT readout in Ref.~\cite{Sailer:2012ua} is not observed in the NaI(Tl)-SiPM setup although slightly increased light yield at around 120\,K is presented in Fig.~\ref{fig:npe}. 
A multiple diffierent effects may be involed in this deviation including characteristics of different NaI(Tl) crystals, integration time window of each analysis~\cite{IANAKIEV2009432}, SiPM's PDE depending on the temperature, and optical interfaces affected by temperature. 
Further study including PMT measurements and different crystals will help to understand this issue. 

\begin{figure}[htbp]
\centering % \begin{center}/\end{center} takes some additional vertical space
\begin{tabular}{cc}
\includegraphics[width=.5\textwidth]{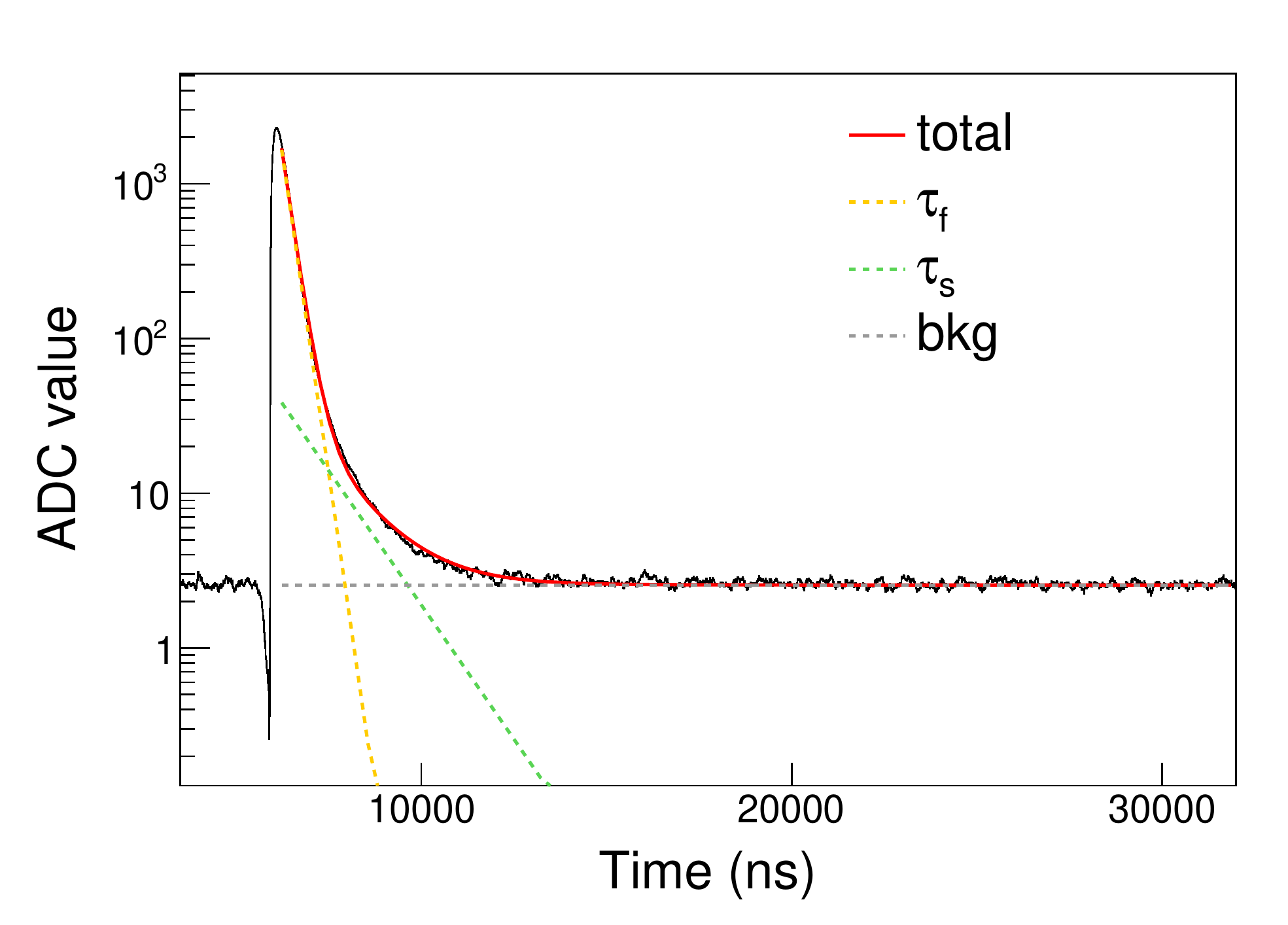} & 
%\qquad
\includegraphics[width=.5\textwidth]{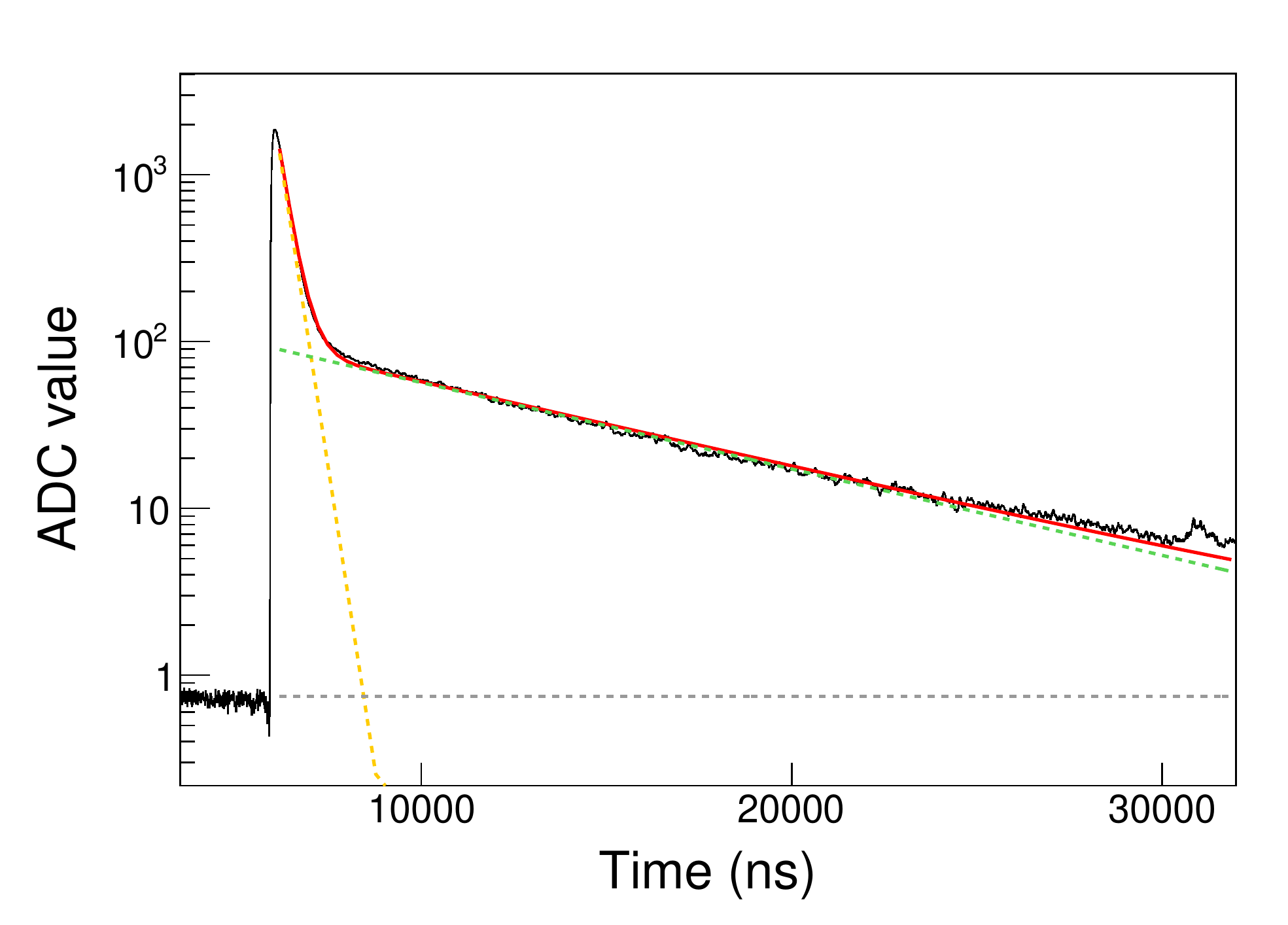} \\
(a) 293 K & (b) 213 K \\
% "\includegraphics" from the "graphicx" permits to crop (trim+clip)
% and rotate (angle) and image (and much more)
\includegraphics[width=.5\textwidth]{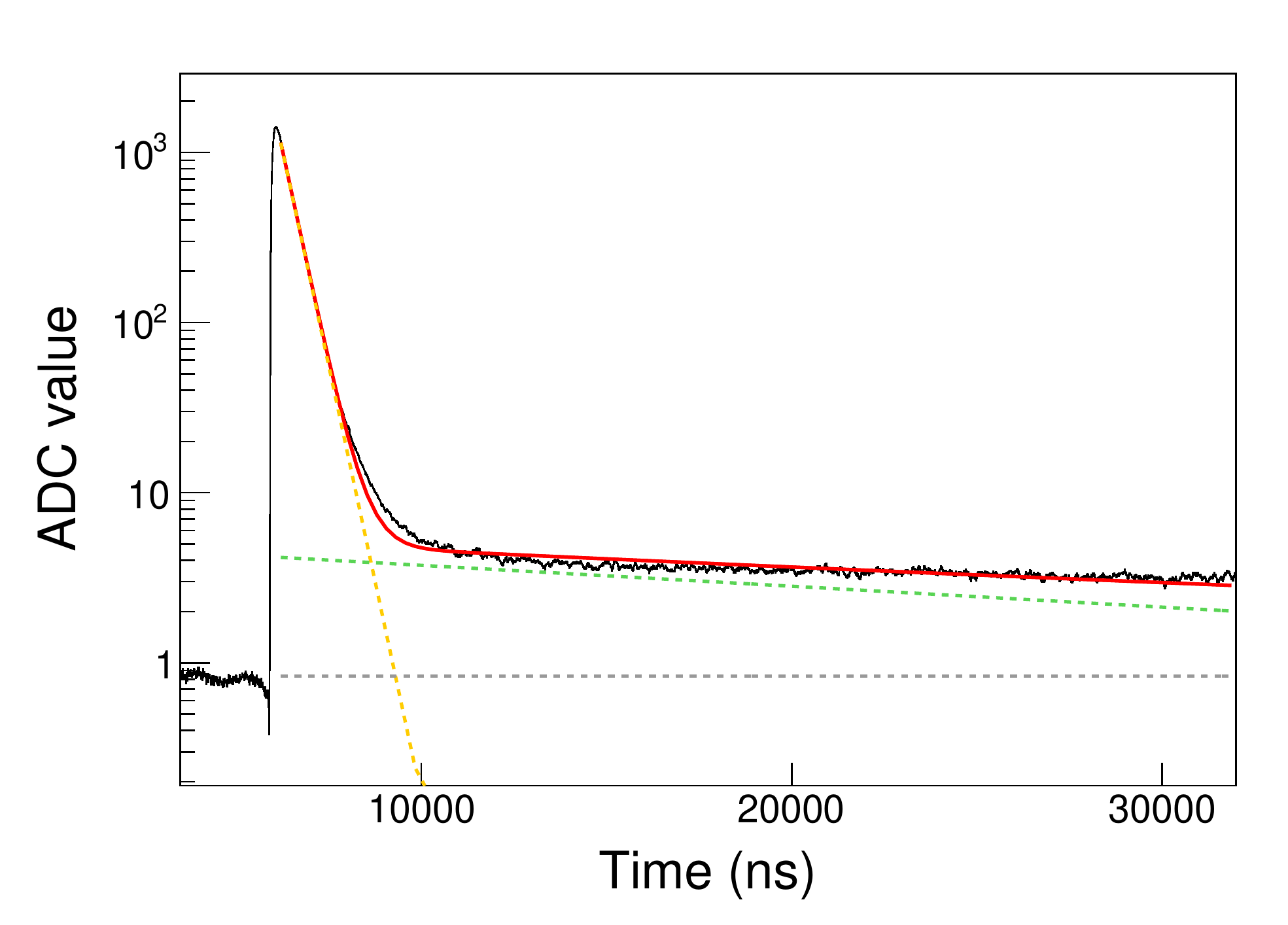} &
\includegraphics[width=.5\textwidth]{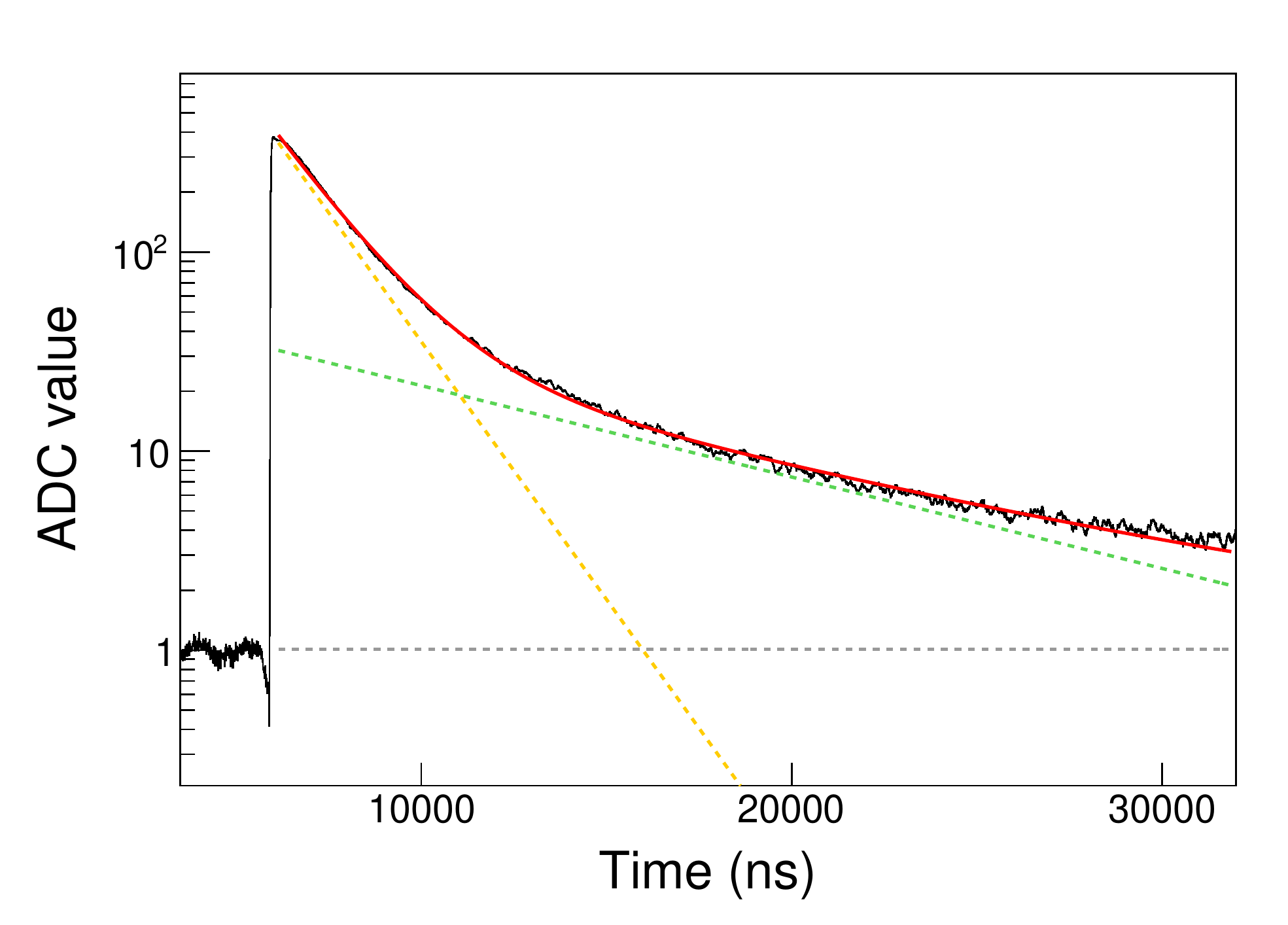} \\
(c) 173 K & (d) 113 K \\
\end{tabular}
\caption{\label{fig:wav}The accumulated waveforms of 59.54 keV gamma rays for four selected temperatures are presented. The red solid line shows the fitted results using two exponentials and a constant background. Each exponential and the constant background are separately overlaided in the plot. }
\end{figure}

We have studied scintillation characteristics of the NaI(Tl) crystal as a function of temperature.
Increased decay time of the NaI(Tl) crystal at low temperatures was already reported~\cite{IANAKIEV2009432,Sailer:2012ua}.
Accumulated waveforms in different temperature measurements are modeled by two exponential functions as follows,
\begin{equation}
\label{eq:1}
F(t) = b_{1}\tau_f \exp^{-(t-t_{0})/{\tau}_{f}} + b_{2}\tau_s \exp^{-(t-t_{0})/{\tau}_{s}} + bkg,
\end{equation}
where ${\tau_{f}}$ and ${\tau_{s}}$ are decay constants for fast and slow decaying components, and ${t_{0}}$ is the rising edge position.
$b_{1}$ and $b_{2}$ are normalization factors and $bkg$ represents continum background from DCR.  
%It has two decay components together with a background pedestal measured in the region 2--4 $\mu$s before the triggered position, around 6 $\mu$s.
Figure~\ref{fig:wav} shows the accumulated waveforms of 59.54 keV gamma events for four selected temperatures of 293, 213, 173, and 113\,K. Each waveform is fitted with Eq.~\ref{eq:1} and fit results are overlaided. 
%At some temperatures, the waveform is not well fitted with two components and looks to be included additional decaying component.
%However, since the fraction of the slow components is not changing even if there is a third component, we present the result considered having two decaying parts here.

% and 1033.00$\pm$37.45 ns
%The previous measurement of the decay constant of NaI(Tl) with PMT at room temperature has fast ($\tau_{f}$) component with a value of 220 ns~\cite{Kim:2014toa}.

\begin{figure}[htbp]
\centering % \begin{center}/\end{center} takes some additional vertical space
\begin{tabular}{cc}
\includegraphics[width=.5\textwidth]{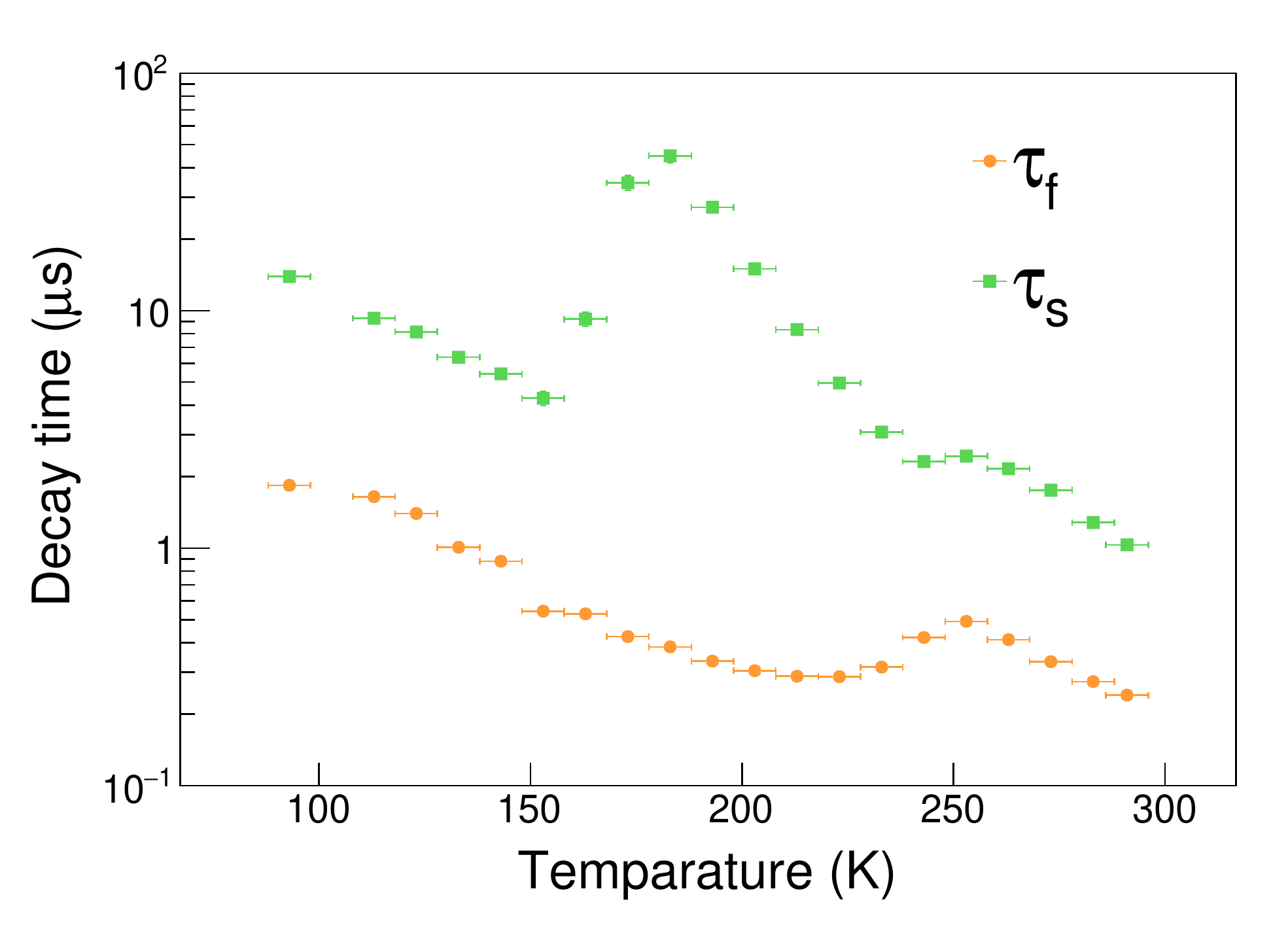} &
\includegraphics[width=.5\textwidth]{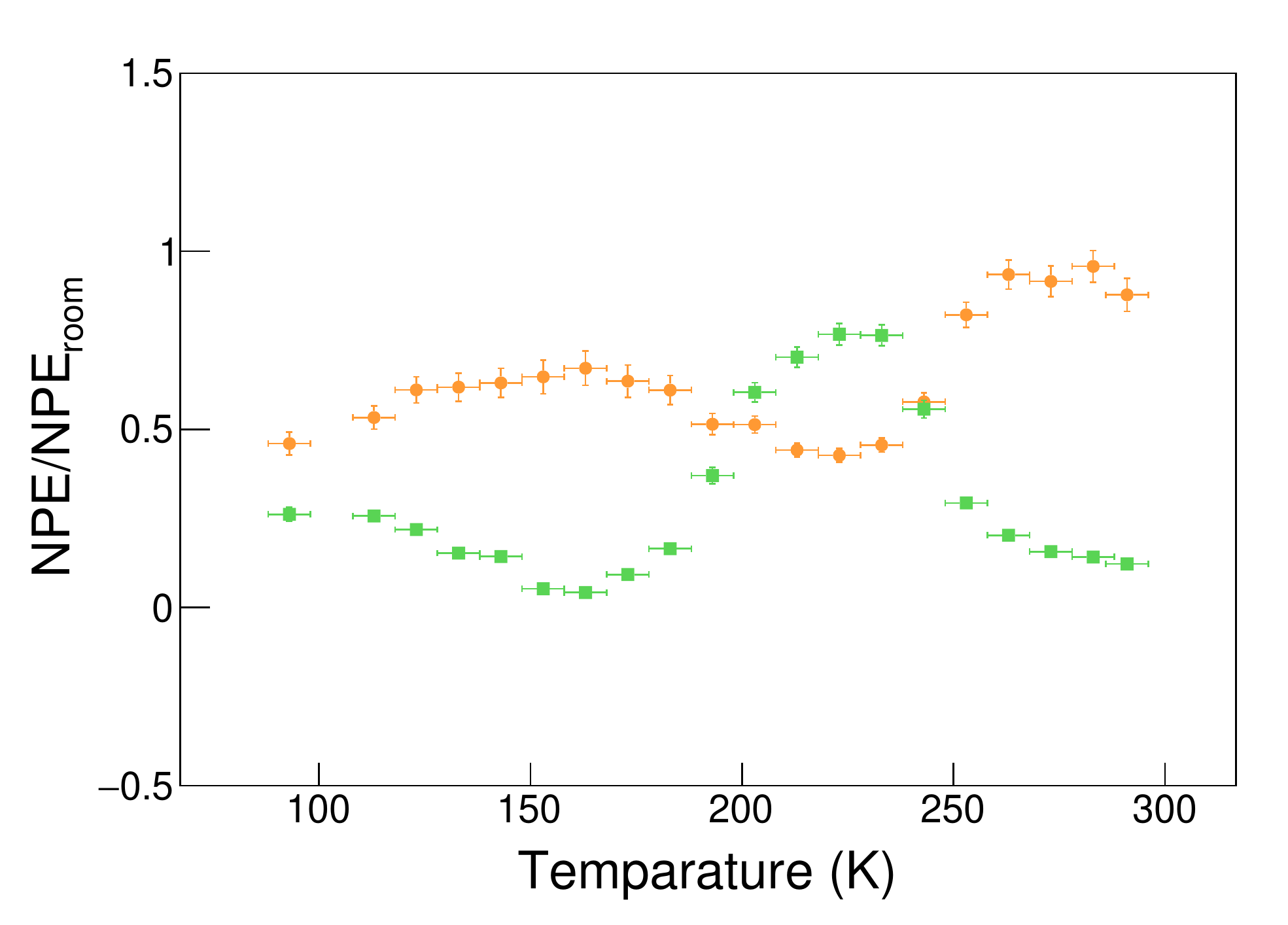} \\
(a) & (b) \\
\end{tabular}
% "\includegraphics" from the "graphicx" permits to crop (trim+clip)
% and rotate (angle) and image (and much more)
\caption{\label{fig:taus} Decay time (a) and the fraction in the signal (b) of fast (yellow circle, $\tau_{f}$), slow (green rectangular, $\tau_{s}$) decay time in different temperatures.}
\end{figure}

The fast decay component at room temperature is approximately 240\,ns that is consistent with previous measurements of about 220\,ns~\cite{Sailer:2012ua,Kim:2014toa}. Slightly longer decay constants is expected due to about 100\,ns width of the SPE pulse from SiPM. 
The decay constants for the fast component ($\tau_f$) and the slow component ($\tau_s$) are measured for a temperature range 93 $-$ 300\,K as shown in Fig.~\ref{fig:taus} (a). The relative amounts of the fast and slow components are shown in Fig.~\ref{fig:taus} (b). 

As shown in Figs.~\ref{fig:wav} and ~\ref{fig:taus}, $\tau_s$ is getting slower below the room temperature down to 170\,K. %Below this temperature, $\tau_s$ is ge170\,K has different behavior like $\tau_f$  slower while $\tau_s$ faster. 
%The fast component is about 90\% of the total charge at the room temperature similar to previous measurements~\cite{IANAKIEV2009432,Kim:2014toa}.
The relative ratio of the fast decay components are 87.8\,$\pm$\,4.6\%, 38.6\,$\pm$\,3.2\%, 87.3\,$\pm$\,4.2\%, and 67.5\,$\pm$\,1.9\% for the temperatures of 293, 213, 173, and 113\,K, respectively. The room temperature measurement with SiPM is  consistent with previous measurements using PMTs~\cite{IANAKIEV2009432,Kim:2014toa}.
The reversed ratios of the fast and slow components around 240\,K were observed in Ref.~\cite{Moszynski2006}.
%{\color{red} let's get the relative ration of the fast components for 293, 213, 173, and 113 K}. 
%291: 0.87 +/- 0.05 (87.77 +/- 4.62)
%213: 0.57 +/- 0.05 (38.61 +/- 3.18)
%173: 0.83 +/- 0.06 (87.28 +/- 4.24)
%113: 0.70 +/- 0.05 (67.49 +/- 1.89)

Maximum increase of the slow component at 230\,K is observed as shown in Fig.~\ref{fig:taus} (b). Interestingly, this temperature has the maximum light yield of the NaI(Tl)-SiPM measurements as shown in Fig.~\ref{fig:npe}. 
The two decay components correspond to the two different scintillation processes. The fast components are caused by the prompt capture of self-trapped excitons by the Tl level~\cite{IANAKIEV2009432,Dietrich19735894}. 
%The population of the self-trapped excitons depends on the temperature. 
Two competing processes to reach the activation centers of exciton by hopping and binary diffusion may cause the slow components~\cite{IANAKIEV2009432,Dietrich19735894}. 
These process are highly depending on the temperature and therefore temperature dependent light yield of the NaI(Tl) crystals~\cite{IANAKIEV2009432,Dietrich19735894,Moszynski2006}.

\begin{figure}[!htbp]
\centering
\includegraphics[width=.8\textwidth]{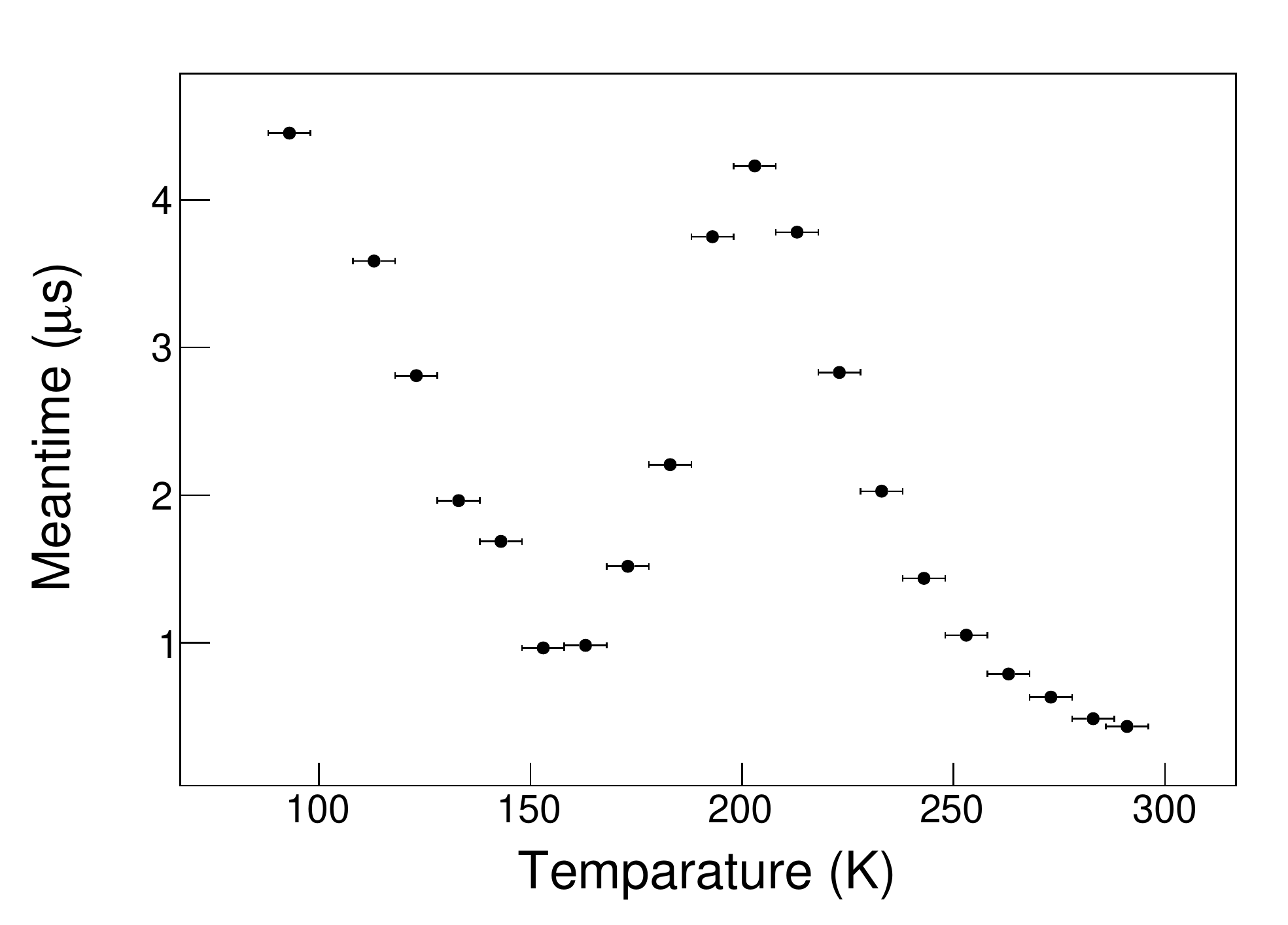}
\caption{\label{fig:mt} Mean time as a function of the temperature. }
\end{figure}

%{\color{red} Let's provide mean decay time with 32$\mu$s windows as another plot}
%The decay characteristics of the scintillation signal, represented with fast and slow decaying components above, can also be described using charge averaged mean decay time defined as below,
We evaluate mean decay times of scintillation pulses in different temperatures defined as follows,
\begin{equation}
\label{eq:2}
\mathrm{Meantime} \equiv \frac{\sum_{i=0}^n h_{i}t_{i}}{\sum_{i=0}^n h_{i}} -t_{0},
\end{equation}
where $h_{i} $ and $t_{i}$ is the height and time of the $i_{th}$ pulse, and $t_{0}$ is the time of the first pulse above the threshold.
The meantime variables were open used to discriminate  nulear recoil events, which can be induced by dark matter interactions, from electron recoil backgrounds~\cite{Lee:2015iaa,KIMS:2018hch}. 
Figure~\ref{fig:mt} shows the meantime as a function of temperature. Increased decay time from 0.3 to 4\,$\mu$s is observed from room temperature to 210\,K. Below this temperature, decreased decay time as low as 1\,$\mu$s until 160\,K is  observed. 
This behavior may related with the observed light yields in Fig.~\ref{fig:npe}. Increased light yields at around 220--250\,K and decreased light yields at around 150--170\,K are indeed correlated with amounts of the long-decay components as shown in Fig.~\ref{fig:taus}. 
Because different nuclear recoil and electron recoil result in different ratios of the fast and slow decay components from different scintillation processes~\cite{knoll}, the increased rate of the slow component from the electron recoil events at around 220\,K may indicate improved pulse shape discrimination for the nuclear recoil events. Further investigation of the pulse shape discrimination using neutron irradiation at the low temperature is desired. 
%This makes the differences in the shape of the signal more pronounced quantitatively.
%The pulse shape discrimination between the different particles is possible due to different scintillation process
%When different particles interact with the material, they undergo other scintillation processes, resulting in a different ratio between fast and slow decay components~\cite{knoll}. 
%Therefore, the interacting particle can be deduced by taking advantage of the meantime difference. 
%The discrimination of particles with the pulse shape is one of the ways to search for dark matter~\cite{Lee:2015iaa,KIMS:2018hch}.
%We have checked the meantime changes of the gamma events according to the temperature variation as shown in Fig.~\ref{fig:mt}.
%It shows a similar tendency to NPE change (Fig.~\ref{fig:npe}), with maximum values at around 210 K and a minimum value at 170 K. 
%As NPE and a meantime are getting larger, there is more chance of good discrimination between different interacting particles.
%It is expected to give better sensitivity for dark matter search if the detector operates at that temperature.
%It will be investigated in the following study to get the quantitative results.

\section{Conclusions}
\label{sec:con}
We have investigated characteristics of a few SiPMs at various temperature points from 300 down to 93\,K. 
Significant decrease of dark count rate at low temperatures indicates that SiPM is a good photosensor for the dark matter search experiment at low temperatures. 
A NaI(Tl)-SiPM detector is also measured with the same temperature range to characterize the NaI(Tl) crystal's light yields and decay times. 
%The DCR is decreased as the temperature getting lower, reaching 0.2 Hz at 140\,K.
The light yield increased  from 300  to 230\,K with a maximal 20$\%$ relative to that of room temperature. 
In a similar temperature range, the ratio of the slow decay component is significantly increased. 
Considering increased light yield and decreased DCR, the NaI(Tl)-SiPM detector can be a good candidate detector for the future dark matter search experiment at low-temperatures. 
%The increased light yield around 150\,K, which was presented by the PMT readout measurement, is not observed with the NaI(Tl)-SiPM setup. 
%We found that the fast and slow decaying time has an increasing tendency as temperature down, and slow components accommodate a substantial fraction of the signal at 250 to 200\,K, corresponding to the region having a high light yield.
%The charge averaged mean decay time also has a similar change to the light yield having a maximum of around 250 to 200\,K.
%The detector is expected to have high discrimination power between particles due to increased light yield and long mean decay time at low temperatures, which makes it to be sensitive for a dark matter search.

\acknowledgments
This work is supported by the Institute for Basic Science (IBS) under project code IBS-R016-A1.

\bibliographystyle{JHEP}
\bibliography{dm}
\end{document}